\documentstyle[preprint,eqsecnum,aps]{revtex}

\begin{document}

\draft

\title{Path Integrals over Velocities in Quantum Mechanics }

\author{Dmitri M. Gitman and Shmaryu M. Shvartsman}

\address{Instituto de F\'{\i}sica, Universidade de S\~ao Paulo \\
Caixa Postal 20516-CEP 01498-970-S\~ao Paulo, S.P., Brasil\\
Department of Physics, Case Western Reserve University \\
Cleveland, OH 44106, USA}

\date{\today}

\maketitle

\begin{abstract}
Representations of propagators  by  means of path integrals over velocities
are discussed both in nonrelativistic and relativistic quantum mechanics.
It is shown that all the propagators can only be expressed through bosonic
path integrals over velocities of space-time coordinates. For spinning and
isospinning particles that is quite
nontrivial statement,  to prove which one needs to do  all
grassmannian integrations in conventional path integral representations.
In the representations the integration over velocities
is not restricted by any boundary conditions; matrices, which have
to be inverted in course of doing Gaussian integrals, do not contain any
derivatives in time, and spinor and isospinor structures of the propagators are
given explicitly. One can define universal Gaussian and quasi-Gaussian
integrals over velocities and rules of handling them.
Such a technique allows one effectively  calculate  propagators in
external fields. Thus, Klein-Gordon propagator is found in  a constant
homogeneous electromagnetic field and its combination with a plane wave field.
\end{abstract}

\section{Introduction}
Propagators of relativistic particles in external fields (electromagnetic,
non-Abelian or gravitational) contain important information about quantum
behavior of these particles. Moreover, if such propagators are
known in an arbitrary external field, one can find exact
one-particle Green's functions in the corresponding quantum field theory,
taking functional integrals over the external field.
It is known that the propagators  can be presented by means of  path integrals
over classical trajectories. Such representations were already
discussed in the literature for a long time in different contexts
\cite{b1,b2,b3,b4,b5,b6,b7,b8,b9,b10,b11,b12,b13,b14,b15,b16}.
Over recent years this activity got some additional motivation to learn on
these simple examples how to quantize by means of path integrals more
complicated theories such as string theory, gravity and
so on. Path integral representations can be effectively used for
calculations of propagators, for example, for concrete calculations of
propagators in external electromagnetic or gravitational fields.
However, in contrast with the field theory, where path integration rules are
enough  well defined, at least in the frame of perturbation theory
\cite{b17,b18}, in relativistic and nonrelativistic quantum mechanics there
are  problems
with uniqueness of definition of path integrals, with boundary conditions,
and so on \cite{b1,b19,b20,b21,b22,b23}.

In this paper we discuss representations of relativistic and nonrelativistic
propagators  by  means of path integrals over velocities. It is shown that
all the propagators can only be expressed through bosonic
path integrals over velocities of space-time coordinates. For spinning and
isospinning particles that is not only a question of convenience, but a
nontrivial statement, to prove which one needs, particularly, to do all
grassmannian integrations in conventional path integral representations
and find spinor and isospinor structure of the propagator explicitly.
Conveniences of the representations are:  the integration over
velocities is not restricted by any boundary conditions, matrices, which have
to be inverted in course of doing Gaussian integrals, do not contain any
derivatives in time, and spinor and isospinor structures of the propagators are
given explicitly by their decomposition in the independent
$\gamma$-matrix structures or generators of a gauge group. One can define
universal Gaussian and quasi-Gaussian
integrals over velocities and rules of handling them. This approach is
similar to one used in the field theory (in the frame of perturbation theory
\cite{b17,b18}). Using such a technique one can effectively  calculate
propagators in external fields. As examples, Klein-Gordon
propagator is found in  a constant
homogeneous electromagnetic field and its combination with a plane wave field.
One ought to say that path integration methods were already
applied for such kind of calculations. For example, the causal propagators
of relativistic
particles in external electromagnetic field of a plane wave were found by
means of path integrations in \cite{b2,b4,b24} and in crossed electric and
magnetic fields in \cite{b5}. More complicated combination of electromagnetic
field, consisting of parallel magnetic and electric field together with a
plane wave, propagating along, was considered in \cite{b3,b12}. In \cite{b8}
they did particular functional integrations to prove a path integral
representation for the causal propagator of spinning particle in an
external electromagnetic field.

\section{Path integrals over velocities in nonrelativistic quantum
mechanics}

In nonrelativistic quantum mechanics they usually consider
path integral representations for the propagation function (amplitude
of the probability) $G(x,t;x't'), \; x=(x^{i}, \; i=\overline{1,3})$,
\begin{eqnarray}\label{e1}
&&G(x,t;x',t')=<x|e^{-i\hat{H}(t-t')}|x'> \;, \\
&&\left(i\frac{\partial}{\partial t}-\hat{H}\right)G(x,t;x',t')=0 \;,
\; \; G(x,t;x',t)=\delta^{3}(x-x') \; .    \nonumber
\end{eqnarray}
\noindent If we suppose, for example, that quantum Hamiltonian $\hat{H}$
is constructed from the classical one $H(x,p)$ by means of Weyl's
ordering procedure, then the following path integral representation takes
place:
\begin{eqnarray}\label{e2}
&&G=G(x_{out},t_{out};x_{in},t_{in})=\int^{x_{out}}_{x_{in}}D'x\int Dp\:
\exp\left\{iS_{H}[x,p]\right\} \nonumber \\
&&=\lim_{N\rightarrow \infty}\int\frac{{\rm d}^{3}p_{N}}{(2\pi)^{3}}
\prod_{k=1}^{N-1}\frac{{\rm d}^{3}x_{k}{\rm d}^{3}p_{k}}{(2\pi)^{3}}
\exp\left\{i\sum_{j=1}^{N}\left[p_{j}\frac{\Delta
x_{j}}{\Delta t}-H(\bar{x}_{j},p_{j})\right]\Delta t\right\}\; ,  \\
&&\Delta x_{j}=x_{j}-x_{j-1}, \;  \Delta t=\frac{t_{out}-t_{in}}{N},
\;  \bar{x}_{j}=\frac{x_{j}+x_{j-1}}{2} \; ,  \nonumber \\
&&S_{H}[x,p]=\int_{t_{in}}^{t_{out}}\left[p\dot{x}-H(x,p)\right]{\rm d}t \; ,
\nonumber
\end{eqnarray}
\noindent where $S_{H}$ is  hamiltonian action and the integration in the
right side of (\ref{e2}) is going over trajectories $x(t)$ with the boundary
conditions $x(t_{in})=x_{in} \; , \; \; x(t_{out})=x_{out}$, and over
trajectories $p(t)$ without any restrictions. We denoted
the functional differential of $x$ with prime to underline the number
of integrations over $x$ is less then one over $p$.

The expression (\ref{e2}) presents a hamiltonian form of the path integral
for propagation function (\ref{e1}). To get a lagrangian form one can make
a shift
\begin{equation}\label{e3}
p\rightarrow p+p_{0} \;,
\end{equation}
\noindent where $p_{0}=p_{0}(x,\dot{x})$ is a solution
of the equation
\[
\frac{\delta S_{H}}{\delta p}=0 \;\Longleftrightarrow \;
\dot{x}=\{x,H\}=\frac{\partial H}{\partial p} \; ,
\]
\noindent with respect to $p$. If $H$ is constructed from a Lagrangian
$L(x,\dot{x})$, then  $p_{0}=\partial L/\partial \dot{x}$, so that
\[
S_{L}[x]=S_{H}[x,p_{0}]=\int_{t_{in}}^{t_{out}}L(x,\dot{x})dt \; .
\]
\noindent and
\[
S_{H}[x,p+p_{0}]=S_{L}[x]+\Delta S_{H} ,\;\; \Delta S_{H}=
\left.-\int_{t_{in}}^{t_{out}}\sum_{n=2}\frac{p^{n}}{n!}\frac
{\partial^{n}H}{\partial p^{n}}\right|_{p=p_{0}}dt \; .
\]
\noindent Thus, one can write the path integral (\ref{e2}) in the following
form
\begin{equation}\label{e4}
G=\int_{x_{in}}^{x_{out}}D'x \;\exp\left\{iS_{L}[x]\right\}
{\cal M}[x]  \; ,
\end{equation}
\noindent with the measure
\begin{equation}\label{e5}
{\cal M}[x]=\int Dp \;\exp\left\{i\Delta S_{H}[x,p]\right\} \; .
\end{equation}
\noindent The expression (\ref{e4}) presents a lagrangian form of the path
integral for the propagation function (\ref{e1}).

One can express the propagation function by means of a path integral over
coordinates and velocities. To this end we make a change of variables in
(\ref{e2}), $(x,p)\rightarrow (x,v)$, where $v$ and $p$ are connected by the
equation
\[
p=\left.\frac{\partial L}{\partial \dot{x}}\right|_{\dot{x}=v}=p_{0}(x ,v) \; .
\]
\noindent (We suppose for simplicity that  Hessian is not zero in the case
of consideration).

\noindent The Jacobian of the change of variables is
\[
J(x,v)={\rm Det}\frac{\partial^{2}L(x,v)}{\partial v^{i}(t)\partial v^{j}
(t)}\delta(t-t') \; ,
\]
\noindent and
\[
S_{H}[x,p_{0}(x,v)]=\int_{t_{in}}^{t_{out}}\left[L(x,v)+\frac{\partial L(x,v)}
{\partial v}(\dot{x}-v)\right]dt \; ,
\]
\noindent so, we get
\begin{equation}\label{e6}
G =\int_{x_{in}}^{x_{out}}D'x\int Dv
\exp\left\{i\int_{t_{in}}^{t_{out}}
\left[L(x,v)+\frac{\partial L(x,v)}{\partial v}(\dot{x}-v)\right]
dt\right\}J(x,v) \; .
\end{equation}
\noindent This formula presents the propagation function by means of
a path integral over coordinates and velocities. A similar formula can be
derived in the field theory for the generating functional of Green's
function \cite{b25}.

Making the shift of velocities, $v\rightarrow v+\dot{x}$, we get again the
lagrangian form (\ref{e4}), but the expression for the measure ${\cal M}[x]$ is
given now
in terms of a path integral over velocities,
\begin{eqnarray}\label{e7}
&&{\cal M}[x]=\int Dv\exp\left\{i\Delta S_{L}[x,v]\right\}
J(x,v)\; , \\
&&\Delta S_{L}[x,v]=-\sum_{n=2}\frac{n-1}{n!}\int_{t_{in}}^{t_{out}}
\frac{\partial^{n}L}{\partial\dot{x}^{n}}v^{n}dt \; . \nonumber
\end{eqnarray}

In case if
\begin{eqnarray}\label{e8}
&&L=L_{0}+L_{int} \; ,\; \; \; L_{0}=\frac{m\dot{x}^{2}}{2} \;, \; \; \;
L_{int}=-V(x) \; ,\\
&&\mbox{or} \nonumber \\
&&H=H_{0}+H_{int} \; , \; \; \; H_{0}=\frac{p^{2}}{2m}\; , \;\; \;
H_{int}=V(x) \; . \nonumber
\end{eqnarray}
\noindent we get simple Feynman's \cite{b1} answer (\ref{e4}) with
$x$-independent measure (\ref{e5}) or (\ref{e7}).

Let us consider different kind of representations, containing path integrals
over velocities. To this end it is useful first to make the number of
integrations over $x$ and $p$ equal in the initial definition (\ref{e2}).
Namely, one can write
\begin{eqnarray}\label{e9}
&&G=\int_{x_{in}}Dx\int Dp \:\delta^{3}(x(t_{out})-x_{out})\exp\left\{iS_{H}
[x,p]\right\} \nonumber \\
&&=\lim_{N\rightarrow \infty}\int\prod_{k=1}^{N}\frac
{{\rm d}^{3}x_{k}{\rm d}^{3}p_{k}}{(2\pi)^{3}}\delta^{3}
(x_{N}-x_{out})\exp\left\{i
\sum_{j=1}^{N}\left[p_{j}\frac{\Delta x_{j}}{\Delta t}-H(\bar{x}_{j},p_{j})
\right]\Delta t\right\}\;,
\end{eqnarray}
\noindent where now only one boundary condition remains, $x(t_{in})=x_{in}$.
Making the shift of momenta (\ref{e3}) in the integral (\ref{e9}), a change of
parameterization of trajectories, introducing instead of time $t$ a parameter
$\tau, \; \; \tau\in[0,1]$,
\[
\tau=\frac{t-t_{in}}{\Delta T} \; , \;\;\; \Delta T=t_{out}-t_{in} \; ,
\]
\noindent replacements
\[
a(x-x_{in}-\tau\Delta x)\rightarrow x,\;\;\;
ap\rightarrow p, \; \; \; \Delta x=x_{out}-x_{in},\;\;\;
a=\sqrt{\frac{m}{\Delta T}} \; ,
\]
\noindent and restricting  ourselves for simplicity with the case (\ref{e8}),
we get the expression\footnote{Here and in what follow we use the
notation $\int d\tau =\int_{0}^{1}d\tau.$}
\begin{eqnarray}\label{e10}
&&G=a^{3}\exp\left\{\frac{im\Delta x^{2}}
{2\Delta T}\right\}\int_{0}Dx\;{\cal M}\;\delta^{3}(x(1)) \nonumber \\
&&\times\exp\left\{i\int{\rm d}\tau\left[\frac{\dot{x}^{2}}{2}
-V(ax+x_{in}+\tau\Delta x)\Delta T
\right]\right\} \; ,
\end{eqnarray}
\noindent where the integration over $x$ is subjected the boundary
condition $x(0)=0$, and the measure ${\cal M}$ has the form
\[
{\cal M}=\int\exp\left\{-\frac{i}{2}\int p^{2}d\tau\right\} \; .
\]

On this step we replace the integration over the trajectories
$x(\tau)$ by one over velocities $v(\tau)$,
\begin{eqnarray}\label{e11}
x( \tau)&=&\int \theta( \tau -\tau
^{\prime })v( \tau ^{\prime })d\tau ^{\prime }=\int_{0}^{\tau}
v( \tau ^{\prime })d\tau ^{\prime }\;, \nonumber \\
v( \tau )&=& \dot {x}( \tau )\;, \;\;\; x(1)=\int vd\tau \; .
\end{eqnarray}
\noindent The corresponding Jacobian can be written as
\[
J={\rm Det}\;\theta ( \tau -\tau ^{\prime })
\]
\noindent and regularized, for example, in the frame of discretization
procedure. Thus, we get
\begin{eqnarray}\label{e12}
&&G=a^{3}\exp\left\{\frac{im\Delta x^{2}}
{2\Delta T}\right\}\int Dv\;{\cal M}\;J\;\delta^{3}\left(\int vd\tau
\right) \\
&&\times\exp\left\{i\int d\tau\left[\frac{v^{2}}{2}
-V(a\int_{0}^{\tau}v(\tau')
d\tau'+x_{in}+\tau\Delta x)\Delta T
\right]\right\} \; , \nonumber
\end{eqnarray}
\noindent where integration over $v$ as well as over $p$ is already not
subjected any boundary conditions.

\noindent One can formally find the Jacobian $J$, switching off the potential
$V(x)$ in (\ref{e11}) and using the expression for the free propagation
function \cite{b1},
\[
G_{0}=\left(\frac{m}{2\pi i\Delta T}\right)^{\frac{3}{2}}
\exp\left\{\frac{im\Delta x^{2}}{2\Delta T}\right\} \; .
\]
\noindent So,
\[
J=\left(\frac {1}{2\pi i}\right)^{\frac{3}{2}}\left[\int Dv\;{\cal M}\;
\delta^3\left( \int v d\tau \right) \exp \left\{ i\int d\tau
\left(\frac{v^2}2\right) \right\} \right] ^{-1} .
\]
\noindent Gathering these results, we may write
\begin{equation}\label{e13}
G=G_{0}\int{\cal D}v\;\delta^{3}\left(\int vd\tau\right)\exp\left\{ i\int d
\tau \left[\frac{v^{2}}{2}-V(a\int
_{0}^{\tau}v(\tau')d\tau'+x_{in}+\tau\Delta x)\Delta T \right]\right\}\; ,
\end{equation}
\noindent where new measure ${\cal D}v$ has the form
\begin{equation}\label{e14}
{\cal D}v=Dv\left[ \int Dv\;\;\delta ^3\left( \int v d\tau
 \right) \exp \left \{i\int d\tau \left(\frac{v^2}2\right)
\right \} \right] ^{-1}\;.
\end{equation}

Thus, we got a representation for the propagation function (\ref{e1}) by
means of a  special kind path integral  over velocities. The conveniences of
this representation are:
 the integration over velocities is not subjected any
boundary conditions and no time derivatives appear in integrand, so, e.g.
matrices, which have to be inverted in course of doing Gaussian integrals,
do not contain any time derivatives. The same kind of path integrals arises in
representations of relativistic particle propagators, which we present in the
next section. One can formulate universal rules of handling such integrals in
the frame  of perturbation theory, what will be done in Sect.4.

\section{Path integrals over velocities in relativistic quantum
mechanics}

\subsection{Scalar particle propagator in an external electromagnetic
field}

As known, the propagator of a scalar particle in an external electromagnetic
field $A_{\mu}(x)$ is the causal Green's function $D^{c}(x,y)$ of the
Klein-Gordon equation in this field,
\begin{equation}\label{f1}
\left[\left( i\partial  -gA\right)^2-m^2+i\epsilon \right]
D^{c}( x,y)=-\delta ^4(x-y)\;,
\end{equation}
\noindent where $x=\left(x^\mu,\; \mu=\overline{0,3} \right)$,  Minkowski
tensor $\eta _{\mu \nu }={\rm diag}(1-1-1-1)$, and infinitesimal term
$i\epsilon$ selects the causal solution.

Consider a lagrangian form of the path integral representation for
$D^{c}\left( x,y\right)$ \cite{b13}, modified by inserting of a $\delta$-
function, similar to the nonrelativistic case,
\begin{eqnarray}\label{f2}
& &D^{c}=D^{c}\left( x_{out},x_{in}\right) =\frac {i}{2}\int_{0}
^{\infty}de_{0}\int _{e_{0}}De\int D\pi\int_{x_{in}}Dx\;M(e)\;\delta^{4}
(x(1)-x_{out})  \nonumber  \\
& &\times \exp \left\{ i\int d\tau \left[ -\frac{\dot
{x}^2}{2e}-\frac{e}{2}m^{2}-g\dot{x}A(x)+\pi\dot{e}\right] \right\} \; ,
\end{eqnarray}
\noindent where  $x^{\mu}(\tau), \; e(\tau),\; \pi(\tau)$ are trajectories
of integration, parameterized by some parameter $\tau\in[0,1]$, and subjected
the boundary conditions $x(0) =x_{in},   \; \; e(0)=e_{0}$; the measure
$M(e)$ can formally be written as
\begin{equation}\label{f3}
M(e)=\int Dp\;\exp\left\{\frac{i}{2}\int ep^{2}d\tau\right\} \;.
\end{equation}
The propagator $D^{c}$ can be only presented via a path integral over
velocities, in the form similar to (\ref{e13}). First we integrate over
$\pi$ and then use the arisen $\delta$-function $\delta(\dot{e})$ to remove
the functional integration over $e$,
\begin{eqnarray*}
&&D^{c}=\frac {i}{2}\int_{0}
^{\infty}\frac{de_{0}}{e_{0}^{2}}\int_{x_{in}}Dx\;M(e_{0})\;\delta^{4}
(x(1)-x_{out}) \\
&&\times\exp \left\{ i\int d\tau \left[ -\frac{\dot
{x}^2}{2e_{0}}-\frac{e_{0}}{2}m^{2}-g\dot{x}A(x)\right] \right\} .
\end{eqnarray*}
\noindent Then, after  the replacement
\begin{equation}\label{f4}
\frac{x-x_{in}-\tau \Delta x}{\sqrt{e_0}}
\rightarrow x\;,\;\;\;\Delta x=x_{out}-x_{in}\;,
\end{equation}
\noindent  we get the expression
\begin{eqnarray}\label{f5}
&&D^{c}=\frac {i}{2}\int_0^\infty \frac{de_0}{e_0^2}\exp \left[- \frac{i}{2}
\left( e_0m^2+
\frac{ \Delta x^{2}}{e_0}\right)\right] \int_{0}Dx \; M(1)\;
\delta^{4}(x(1)) \nonumber\\
&&\times \exp \left\{ i\int d\tau \left[ -\frac{\dot{x}^{2}}{2}-
g( \sqrt{e_0}\dot x+\Delta x) A( \sqrt{e_0}x+x_{in}+\tau
\Delta x) \right] \right \} \; ,
\end{eqnarray}
\noindent where the trajectories $x$
obey already zero boundary conditions, $x(0)=0$.

As in nonrelativistic case, we replace the integration over the trajectories
$x$ by one over velocities $v$, according to eq.
(\ref{e11}). Thus,
\begin{eqnarray}\label{f6}
&&D^c=\frac {i}{2}\int_0^\infty \frac{de_0}{e_0^2}\exp \left[ -\frac {i}{2}
\left( e_0m^2+ \frac{ \Delta x^2 }{e_0}\right)\right]\int Dv\;M(1)\;J\;\delta ^
{4} \left(\int v d\tau \right)  \\
&&\times\exp \left \{ i\int d\tau \left[ -\frac{v^2}{2}-g\left( \sqrt{e_{0}}v
+\Delta x \right) A\left( \sqrt{e_0}\int_{0}^{\tau} v( \tau ^{\prime })
d\tau ^{\prime }+x_{in}+\tau \Delta x \right ) \right ] \right \}\;.
\nonumber
\end{eqnarray}
\noindent One can formally find the Jacobian $J$, switching off the
potential $A_{\mu} (x)$ in (\ref{f6}) and using the expression
for the free causal Green's function $ D_{0}^{c}$,
\[
D_{0}^{c}=D_{0}^{c}( x_{out},x_{in}) =\frac {1}{2( 2\pi )
^{2}}\int_{0}^{\infty} \frac{de_0}{e_0^2}\exp \left[-\frac {i}{2}
\left( e_0m^2+\frac{\Delta x^2}{e_0} \right) \right]\;.
\]
So, we get formally
\begin{equation}\label{f7}
J=\frac {1}{i\left( 2\pi \right) ^2}\left[ \int Dv\;M(1)\;\delta
^4\left( \int v d\tau \right) \exp \left\{ i\int d\tau
\left( -\frac{v^2}{2}\right) \right\} \right] ^{-1} .
\end{equation}
\noindent Gathering these results, we may write
\begin{eqnarray}\label{f8}
& &D^c=\frac {1}{2(2\pi)^2}\int_{0}^{\infty}
\frac{de_0}{e_0^2}\exp \left[-\frac {i}{2}\left( e_0m^2+\frac{
\Delta x^2 }{e_0} \right) \right] \Delta_{1} ( e_0)\;, \\
& &\Delta_{1} ( e_0)=\int {\cal D} v\;\delta ^4\left( \int v d\tau \right)
\exp \left \{ i\int d\tau \left[ -\frac{v^2}{2}
-g\left( \sqrt{e_{0}}v+\Delta x \right)\right. \right.
\nonumber \\
& &\left.\left. \times A\left( \sqrt{e_0}\int_{0}^{\tau}
v( \tau ^{\prime })d\tau ^{\prime }
+x_{in}+\tau \Delta x \right ) \right] \right \}\;,\label{f9}
\end{eqnarray}
\noindent where the new measure ${\cal D}v$ has the form
\begin{equation}\label{f10}
{\cal D}v=Dv\left[ \int Dv\;\;\delta ^4\left( \int v d\tau
 \right) \exp \left \{ i\int d\tau \left(-\frac{v^2}2\right)\right \}
\right] ^{-1}\;.
\end{equation}
\noindent It is clear that $\Delta_{1}( e_0) =1$ at  $A=0$.

Thus, we got a representation for the propagator (\ref{f1}) by means of a
path integral over velocities of similar kind as in nonrelativistic case.

\subsection{Spinning particle propagator in external electromagnetic field}

The propagator of a spinning particle in an external electromagnetic
field $A_{\mu}(x)$ is the causal Green's function $S^{c}(x,y)$ of the Dirac
equation in this field,
\begin{eqnarray}\label{f11}
&&\left[\gamma^{\mu}\left(i\partial_{\mu}-gA_{\mu}\right)-m\right]
S^{c}(x,y)=-\delta^{4}(x-y)\;,\\
&&\left[\gamma^{\mu}\;,\;\gamma^{\nu}\right]_{+}=2\eta^{\mu \nu}\;.
\nonumber
\end{eqnarray}

Consider a lagrangian form of the path integral representation for,
transformed by $\gamma^{5}=\gamma^{0}\gamma^{1}\gamma^{2}\gamma^{3}$
function  $\tilde{S}^{c}(x,y)=S^{c}(x,y)\gamma^{5}$, (see\cite{b13}),
modified as in the two previous cases by inserting of a $\delta$-function,
\begin{eqnarray}\label{f12}
&&\tilde{S}^{c}=\tilde{S}^{c}(x_{out},x_{in})=\exp\left\{i\gamma^{n}
\frac{\partial_{\ell}}{\partial \theta^{n}}\right\}\int_{0}^{\infty}de_{0}
\int d\chi_{0}\int_{e_{0}}De\int_{\chi_{0}}D\chi \int_{x_{in}}Dx\int D\pi_{e}
\int D\pi_{\chi} \nonumber \\
&&\times\int_{\psi(1)+\psi(0)=\theta} {\cal D}\psi\;
M(e)\;\delta^{4}(x(1)-x_{out})
\exp\left\{i\int\left[-\frac{\dot{x}^{2}}
{2e}-\frac{e}{2}m^{2}-g\dot{x}A(x) \right.\right.\nonumber \\
&&+iegF_{\mu \nu}(x)\psi^{\mu}\psi^{\nu}
\left.\left.\left. +i\left(\frac{\dot{x}_{\mu}\psi^{\mu}}{e}-m\psi^{5}\right)
\chi-i\psi_{n}\dot{\psi}^{n}+\pi_{e}\dot{e}+\pi_{\chi} \dot{\chi}\right]d\tau
+ \psi_{n}(1)\psi^{n}(0)\right\}\right|_{\theta=0}\;,
\end{eqnarray}
\noindent where
\[
\left[\gamma^{m}\;,\;\gamma^{n}\right]_{+}=2\eta^{mn}, \; m,n=\overline
{0,3},5; \; \;
\eta^{mn}={\rm diag}(1-1-1-1-1) \; ;
\]
\noindent $\theta^{n}$ are auxiliary grassmannian (odd) variables,
anticommuting by definition with the $\gamma$-matrices;  $x^{\mu}(\tau), \;
e(\tau), \; \pi_{e}(\tau)$ are bosonic trajectories of integration;
$\psi^{n}(\tau), \; \chi(\tau), \; \pi_{\chi}(\tau)$ are odd
trajectories of integration; and boundary conditions
$x(0)=x_{in}, \;
e(0)=e_{0},\;\psi^{n}(0)+\psi^{n}(1)=\theta^{n}, \;
\chi(0)=\chi_{0} $ take place; the measure $M(e)$ is defined in (\ref{f3})
and
\begin{equation}\label{f13}
{\cal D}\psi=D\psi\left[\int_{\psi(0)+\psi(1)=0} D\psi \exp\left\{\int_{0}^{1}
 \psi_{n}\dot{\psi}^{n}
d\tau \right\}\right]^{-1} \; ,
\end{equation}
\noindent is the measure in the integration over $\psi$.

We are going to demonstrate that the propagator of a spinning particle can
also be expressed  through a bosonic path integral over velocities of
coordinates $x$. To this end one needs to fulfil several functional
integrations. First, one can integrate over $\pi_{e}$ and $\pi_{\chi}$,
and then use arisen $\delta$-functions to remove the functional integration
over $e$ and $\chi$,
\begin{eqnarray}\label{f14}
&&\tilde{S}^{c}=-\exp\left\{i\gamma^{n}
\frac{\partial_{\ell}}{\partial \theta^{n}}\right\}
\int_{0}^{\infty}de_{0}\;\int_{x_{in}}Dx\;
\int_{\psi(0)+\psi(1)=\theta} {\cal D}\psi\;M(e_{0})\;\delta^{4}(x(1)-x_{out})
\nonumber \\
&&\times  \int \left(\frac{\dot{x}_{\mu}\psi^{\mu}}{e_{0}}-
m\psi^{5}\right)d\tau
\exp\left\{i\int\left[-\frac{\dot{x}^{2}}{2e_{0}}-\frac{e_{0}}{2}
m^{2}-g\dot{x}A(x) \right.
\right.\nonumber \\
&&\left.\left.\left.+ige_{0}F_{\mu \nu}(x)\psi^{\mu}\psi^{\nu}-i\psi_{n}
\dot{\psi}^{n}\right]d\tau+ \psi_{n}(1)\psi^{n}(0)\right\}\right|_{\theta=0}\;,
\end{eqnarray}
\noindent Then, it is convenient to replace the integration over $\psi$ by one
over related odd velocities $\omega$,
\begin{equation}\label{f15}
\psi(\tau)=\frac{1}{2}\int \varepsilon (\tau-\tau')\omega(\tau')d\tau'+
\frac{1}{2}\theta\;,\;\;\omega(\tau)=\dot{\psi}(\tau)\;,\;
\varepsilon (\tau)={\rm sign}\;\tau\;.
\end{equation}
\noindent There are not more any restrictions on $\omega$; because of
(\ref{f15}) the boundary conditions for $\psi$ are obeyed automatically.
The corresponding
Jacobian does not depend on variables and cancels with the same one from the
measure (\ref{f13}). Thus\footnote{Here and further
we are using condensed notations,
 $\omega\varepsilon
\omega=\int d\tau d\tau'\omega(\tau)\varepsilon(\tau-\tau')\omega(\tau'),
\; \dot{x}A(x)=\int d\tau\dot{x}A(x)$ and so on.},
\begin{eqnarray}\label{f16}
&&\tilde{S}^{c}=-\frac{1}{2}\exp\left\{i\gamma^{n}
\frac{\partial_{\ell}}{\partial \theta^{n}}\right\}
\int_{0}^{\infty}de_{0}\;\int_{x_{in}}Dx\;
\int {\cal D}\omega\; M(e_{0})\;\delta^{4}(x(1)-x_{out})
\nonumber \\
&&\times \left[\frac{\dot{x}_{\mu}}{e_{0}}\left(\varepsilon \omega^{\mu}+
\theta^{\mu}\right)-
m\left(\varepsilon \omega^{5}+\theta^{5}\right)\right]
\exp\left\{i\left[-\frac{\dot{x}^{2}}{2e_{0}}-\frac{e_{0}}{2}
m^{2}-g\dot{x}A(x) \right.\right.
\nonumber \\
&&\left.\left.\left.-\frac{ie_{0}g}{4}\left(\omega^{\mu}\varepsilon-
\theta^{\mu}\right)F_{\mu \nu}(x)\left(\varepsilon \omega^{\nu}+
\theta^{\nu}\right)+\frac{i}{2}\omega_{n}\varepsilon\omega^{n}
\right]\right\}\right|_{\theta=0}\;,
\end{eqnarray}
\noindent where the measure ${\cal D}\omega$ is
\begin{equation}\label{f17}
{\cal D}\omega=D\omega\left[\int D\omega \exp\left\{-\frac{1}{2}
\omega^{n}\varepsilon
\omega_{n}\right\}\right]^{-1}\;.
\end{equation}
\noindent One can prove, that for a function $f(\theta)$ in the Grassmann
algebra, the following identity holds
\begin{eqnarray}\label{f18}
&&\left.\exp\left\{i\gamma^{n}
\frac{\partial_{\ell}}{\partial \theta^{n}}\right\}f(\theta)\right|
_{\theta=0}=\left.f\left(\frac{\partial_{\ell}}
{\partial \zeta}\right)
\exp\left\{i\zeta_{n}\gamma^{n}\right\}\right|_{\zeta=0}
\nonumber \\
&&=\left.\sum_{k=0}^{4}\sum_{n_{1}\cdots n_{k}}f_{n_{1}\cdots n_{k}}
\frac{\partial_{\ell}}{\partial \zeta_{n_{1}}}\cdots
\frac{\partial_{\ell}}{\partial \zeta_{n_{k}}}
\sum_{l=0}^{4}\frac{i^{l}}{l!}\left(\zeta_{n}\gamma^{n}
\right)^{l}\right|_{\zeta=0}\; ,
\end{eqnarray}
\noindent where $\zeta_{n}$ are some odd
variables. Taking (\ref{f18}) into account in (\ref{f16}), we get
\begin{eqnarray}\label{f19}
&&\tilde{S}^{c}=-\frac{1}{2}\int_{0}^{\infty}de_{0}\;\int_{x_{in}}Dx\;M(e_{0})
\;\delta^{4}\left(x(1)-x_{out}\right)\\
&&\times\left[\frac{\dot{x}_{\mu}}{e_{0}}\left(\varepsilon\frac{\delta_{\ell}}
{\delta \rho_{\mu}}
+\frac{\partial_{\ell}}{\partial \zeta_{\mu}}\right)
-m\left(\varepsilon\frac{\delta}{\delta \rho_{5}}+i\gamma^{5}\right)\right]
\exp\left\{i \left[-\frac{\dot{x}^{2}}{2e_{0}}-\frac{e_{0}}{2}
m^2 \right.\right.
\nonumber \\
&&\left.\left.-g\dot{x}A(x)+\frac{ie_{0}g}{4}F_{\mu \nu}(x)
\frac{\partial_{\ell}}{\partial \zeta_{\mu}}
\frac{\partial_{\ell}}{\partial \zeta_{\nu}}\right] \right\}
\left.R\left[x,\rho,\frac{\partial_{\ell}}{\partial \zeta}\right]
\exp\left\{i\zeta_{\mu}\gamma^{\mu}\right\}\right|_{\rho=0\;,\;\zeta=0}\;,
\nonumber
\end{eqnarray}
\noindent where $\rho_{n}(\tau)$
are  odd sources for $\omega^{n}(\tau)$ and
\begin{eqnarray}\label{f20}
&&R\left[x,\rho,\frac{\partial_{\ell}}{\partial \zeta}\right]
=\int {\cal D}\omega\;\exp\left\{-\frac{1}{2}
\omega^{n}T_{nk}(x|g)\omega^{k}
+I_{n}\omega^{n}\right\}\;,\\
&&I_{\mu}=\rho_{\mu}-\frac{e_{0}g}{2}
\frac{\partial_{\ell}}{\partial \zeta_{\nu}}
F_{\nu \mu}(x)\varepsilon\;,\;\;I_{5}=\rho_{5}\;,
\nonumber
\end{eqnarray}
\noindent with
\begin{equation}\label{f21}
T_{nk}(x|g)=\left ( \begin{array}{cc}
          \Lambda_{\mu \nu}(x|g) & 0\\

          0                               &-\varepsilon\\
        \end{array}\right )
\;,\;\;\Lambda_{\mu \nu}(x|g)=\eta_{\mu \nu}\varepsilon
-\frac{e_{0}}{2} \varepsilon gF_{\mu \nu}(x)
\varepsilon\;.
\end{equation}
Integral in (\ref{f20}) is gaussian one. It can be easily done
\cite{b26}, taking into account its original definition \cite{b13},
\begin{equation}\label{f22}
R\left[x,\rho,\frac{\partial_{\ell}}{\partial \zeta}\right]=
\left[\frac{{\rm Det}T(x|g)}{{\rm Det}T(x|0)}\right]^{1/2}
\exp\left\{-\frac{1}{2}I_{n}
\left[T^{-1}(x|g)\right]^{nk}I_{k}\right\}\;,
\end{equation}

\begin{equation}
\label{f23}
\left[T^{-1}(x|g)\right]^{nk}=\left ( \begin{array}{cc}
          \left(\Lambda^{-1}(x|g)\right)^{\mu \nu} & 0\\

          0                               &-\varepsilon^{-1}\\
        \end{array}\right )
\;.
\end{equation}

The ratio ${\rm Det}T(x|g)/{\rm Det}T(x|0)$ in (\ref{f22}) can be
replaced by ${\rm Det}\Lambda(x|g)/{\rm Det}\Lambda(x|0)$ due to the
structure (\ref{f21}) of the matrix $T(x|g)$, and the latter can be presented
in a convenient form, which allows one to avoid problems with calculations
of determinants of matrices with continuous indices. Namely,
let us differentiate the well known formula
\[
{\rm Det}\Lambda(x|g)=\exp\left\{{\rm Tr}\;\ln \Lambda(x|g)\right\}
\]
\noindent with respect to $g$. So we get the equation
\[
\frac{d}{dg}{\rm Det}\Lambda(x|g)={\rm Det}\Lambda(x|g)\;{\rm Tr}\;\Lambda^{-1}
(x|g)\frac{d\Lambda(x|g)}{dg}=-e_{0}g{\rm Det}\Lambda(x|g){\rm Tr}\;
{\cal G}(x|g) F(x)\;,
\]
\noindent with
\begin{equation}\label{f24}
{\cal G}^{\mu \nu}(x|g)=\frac{1}{2}\varepsilon
\left[\Lambda^{-1}(x|g)\right]^{\mu \nu}\varepsilon\;.
\end{equation}
\noindent This equation can be solved in the form
\begin{equation}\label{f25}
\frac{{\rm Det}\Lambda(x|g)}{{\rm Det}\Lambda(x|0)}
=\exp\left\{-e_{0}\int_{0}^{g}dg'
\;{\rm Tr}\;{\cal G}(x|g') F(x)\right\}\;.
\end{equation}

Besides, the representation (\ref{f19}) contains
only first derivatives with respect to $\rho_{n}(\tau)$, acting on
$R$ at $\rho_{n}=0$. This circumstance allows one to replace in (\ref{f19})
the expression (\ref{f22}) by
\begin{eqnarray}\label{f26}
&&\tilde{R}\left[x,\rho,\frac{\partial_{\ell}}{\partial \zeta}\right]=
\left[\frac{{\rm Det}\Lambda(x|g)}{{\rm Det}\Lambda(x|0)}\right]^{1/2}
\\
&&\times \exp\left\{e_{0}g\rho_{\mu}\varepsilon ^{-1}{\cal G}^{\mu \alpha}(x|g)
F_{\alpha \nu}(x)\frac{\partial_{\ell}}{\partial\zeta_{\nu}}-
\frac{e_{0}^{2}g^{2}}{4}\left(F(x){\cal G}(x|g)
F(x)\right)_{\mu \nu}
\frac{\partial_{\ell}}{\partial\zeta_{\mu}}
\frac{\partial_{\ell}}{\partial\zeta_{\nu}}\right\}\;.
\nonumber
\end{eqnarray}
\noindent Substituting (\ref{f25},\ref{f26}) into (\ref{f19}), and performing
functional differentiation with respect to $\rho_{\mu}$, we get
\begin{eqnarray}
\label{f27}
&&\tilde{S}^{c}=-\frac{1}{2}\int_{0}^{\infty}de_{0}\;\int_{x_{in}}Dx\;M(e_{0})
\;\delta^{4}\left(x(1)-x_{out}\right)\left[\frac{\dot{x}^{\mu}}
{e_{0}}K_{\mu \nu}(x)
\frac{\partial_{\ell}}{\partial \zeta_{\nu}}
-im\gamma^{5}\right]
\nonumber \\
&&\times\exp\left\{i \left[-\frac{\dot{x}^{2}}{2e_{0}}-\frac{e_{0}}{2}
m^2 -g\dot{x}A(x)+\frac{ie_{0}}{2}\int_{0}^{g}dg'
\;{\rm Tr}\;{\cal G}(x|g') F(x)\right.\right.
\nonumber \\
&&\left.\left.\left.+\frac{ie_{0}g}{4}\left(F(x)
K(x)\right)_{\mu \nu}
\frac{\partial_{\ell}}{\partial \zeta_{\mu}}
\frac{\partial_{\ell}}{\partial \zeta_{\nu}}\right] \right\}
\exp\left\{i\zeta_{\mu}\gamma^{\mu}\right\}\right|_{\zeta=0}\;,
\end{eqnarray}
\noindent where
\begin{equation}\label{f28}
K_{\mu \nu}=\eta_{\mu \nu}+e_{0}g\left({\cal G}(x|g)
F(x)\right)_{\mu \nu}\;.
\end{equation}

\noindent The differentiation over $\zeta$ in (\ref{f27}) can be fulfilled
explicitly, using eq.(\ref{f18}),
\begin{eqnarray}\label{f29}
&&S^{c}=\frac{i}{2}\int_{0}^{\infty}de_{0}\;\int_{x_{in}}Dx\;M(e_{0})
\;\delta^{4}\left(x(1)-x_{out}\right)\Phi(x,e_{0})\nonumber    \\
&&\times \exp\left\{i \left[-\frac{\dot{x}^{2}}{2e_{0}}-\frac{e_{0}}{2}
m^2 -g\dot{x}A(x)\right]\right\}\;, \\
&&\Phi(x,e_{0})=\left[m+(2e_{0})^{-1}\dot{x}K(x)\left(1-2gF(x)K(x)\right)
\gamma+im\frac{e_{0}g}{4}\left(F(x)K(x)\right)_{\mu \nu}\sigma^{\mu \nu}
\right. \nonumber \\
&&+\left.i\frac{g}{4}\left(\dot{x}K(x)\gamma\right)\left(F(x)K(x)\right)
_{\mu \nu}\sigma^{\mu \nu}+m\frac{e^{2}_{0}g^{2}}{16}\left(F(x)K(x)\right)_
{\mu \nu}\left(F(x)K(x)\right)^{\ast \mu \nu}\gamma^{5}\right]
\nonumber \\
&&\times\exp\left\{-\frac{e_{0}}{2}\int_{0}^{g}dg'
\;{\rm Tr}\;{\cal G}(x|g') F(x)\right\}\;, \label{f30}
\end{eqnarray}
\noindent where
\begin{equation}\label {f31}
\sigma^{\mu \nu}=\frac{i}{2}\left[\gamma^{\mu}\;,\;\gamma^{\nu}\right]
\;,\;\;\left(F(x)K(x)\right)^{\ast \mu \nu}=\frac{1}{2}\varepsilon^{\mu \nu
\alpha \beta}\left(F(x)K(x)\right)_{\alpha \beta}\;,
\end{equation}
\noindent and $\varepsilon^{\mu \nu \alpha \beta}$ is Levi-Civita symbol.

The eq.(\ref{f30}) gives a representation for Dirac propagator
as a path integral over bosonic
trajectories of a functional, which spinor structure is found
explicitly, namely, its decomposition in all independent
$\gamma$-structures is given. The functional $\Phi(x,e_{0})$ can be called
spin factor, and namely it distinguishes Dirac propagator from the
scalar one. One needs to stress that spin factor is gauge invariant, because
of its dependence of $F_{\mu \nu}(x)$ only.

Making the replacement (\ref{f4}) in (\ref{f29}) and going over to the
integration over velocities $v$, according (\ref{e11}), and using the
expression (\ref{f7}) for the corresponding Jacobian, one can present
Dirac propagator via path integral over velocities only,
\begin{eqnarray}\label{f32}
& &S^c=\frac {1}{2(2\pi)^2}\int_{0}^{\infty}
\frac{de_0}{e_0^2}\exp \left[-\frac {i}{2}\left( e_0m^2+\frac{
\Delta x^2 }{e_0} \right) \right] \Delta_{2} ( e_0)\;, \\
& &\Delta_{2}(e_0)=\int {\cal D} v\;\delta ^4\left( \int v d\tau \right)
\Phi(\sqrt{e_0}\int_{0}^{\tau} v( \tau ^{\prime })d\tau ^{\prime }
+x_{in}+\tau \Delta x,e_{0}) \nonumber \\
&&\exp \left \{ i\int d\tau \left[ -\frac{v^2}{2}
-g(\sqrt{e_{0}}v+\Delta x)A\left( \sqrt{e_0}\int_{0}^{\tau}
v( \tau ^{\prime })d\tau ^{\prime }
+x_{in}+\tau \Delta x \right ) \right] \right \}\;,\label{f33}
\end{eqnarray}
\noindent where the new measure ${\cal D}v$ has the form
(\ref{f10})

\subsection{Scalar particle propagator in non-Abelian external
field}

In the same manner one can present relativistic particle propagators in
non-Abelian external  field. Here we restrict ourselves with a consideration
of a scalar particle propagator in an external electromagnetic $A_{\mu}(x)$
and non-Abelian $B_{\mu}(x)$ fields. Such a propagator is the causal
Green's function $D^{c}(x,y)$ of the Klein-Gordon equation in the fields,
\[
\left[\left(i\partial-gA(x)-B^{a}(x)T_{a}\right)^{2}-m^{2}\right]
D^{c}(x,y)=-\delta^{4}(x-y)\;,
\]
where $T^{a}$ are generators of a corresponding group. Choosing for
simplicity $SU(2)$ as the group, we have $T_{a}=\frac{1}{2}\sigma_{a}$,
where $\sigma_{a}$ are Pauli matrices.
The propagator $D^{c}$ can be presented via bosonic and grassmannian path
integrals \cite{b9,b12},
\begin{eqnarray}\label{f34}
&&D^{c}=D^{c}(x_{out},x_{in})=\frac{i}{2}\exp\left\{i\sigma_{a}
\frac{\partial_{\ell}}{\partial \theta_{a}}\right\}\int_{0}^{\infty}de_{0}
\int_{e_{0}}De \int_{x_{in}}^{x_{out}}Dx\int D\pi_{e}
\int_{\phi(0)+\phi(1)=\theta} {\cal D}\phi \nonumber \\
&&\times M(e)
\left.\exp\left\{i\left[-\frac{\dot{x}^{2}}
{2e}-\frac{e}{2}m^{2}-g\dot{x}A(x)-\dot{x}B^{a}(x){\cal T}_{a}
-i\phi_{a}\dot{\phi}_{a}+\pi_{e} \dot{e}\right]+\phi_{a}(1)\phi_{a}(0)\right\}
\right|_{\theta=0}\;, \\
&&{\cal D}\phi=D\phi \left[\int_{\phi(0)+\phi(1)=0}
 D\phi \exp\left\{\phi_{a}\dot{\phi}_{a}\right\}
\right]^{-1}\;,
\nonumber
\end{eqnarray}
where  $\theta_{a}$ are auxiliary odd variables, anticommuting by definition
with the $\sigma$-matrices; $\phi_{a}(\tau)$ are odd trajectories of
integration and ${\cal T}_{a}=-i\epsilon_{abc}\phi_{b}\phi_{c}$.
All grassmannian integrals can be done
similar to the spinning particle case and final result presented in the form
\begin{eqnarray}\label{f35}
D^{c}&=&\frac{i}{2}\int_{0}^{\infty}de_{0}\int_{x_{in}}^{x_{out}}Dx\;
M(e_{0})\;\Phi(x)
\exp\left\{i\left[-\frac{\dot{x}^{2}}{2e_{0}}-\frac{e_{0}}{2}m^{2}-
g\dot{x}A(x)\right]\right\}\;,\\
\Phi(x)&=&\left[1+{\rm Tr}\;R(x){\cal G}(x|1)R(x)\;-\frac{i}{2}
L_{ab}(x)\epsilon_{abc}T_{c}\right]
\exp\left\{-\frac{1}{2}\int_{0}^{1}d\lambda \;{\rm Tr}\;{\cal G}(x|\lambda)
R(x)\right\}\;,
\label{f36}\\
&&{\cal G}(x|\lambda)=\frac{1}{2}\varepsilon Q^{-1}(x|\lambda)\varepsilon\;,\;
Q(x|\lambda)=\varepsilon I-\frac{\lambda}{2}\varepsilon R(x)
\varepsilon\;,\;R_{ab}(x)=\dot{x}B^{c}(x)\epsilon_{cab}\;,
\nonumber\\
&&L_{ab}(x)=R_{ab}(x)-\left[R(x){\cal G}(x|1)R(x)\right]_{ab}\;,
\nonumber
\end{eqnarray}
where $I$ is unit matrix in the group space.
The isospinor factor (\ref{f36}) in (\ref{f35}) is
presented by its decomposition in the generators $T_{a}$ of the $SU(2)$ group.
Explicit description of the spinor and isospinor structure of Dirac propagator
in both Abelian and non-Abelian external fields is more complicated problem
which, nevertheless, can be solved in the frame of the same approach.

Propagator $D^{c}$ can be written in terms of path integral over velocities
as in spinning particle case. The result has the form (\ref{f32},\ref{f33}),
where spinor factor $\Phi(x,e_{0})$ has to be replaced by the isospinor
factor $\Phi(x)$ (\ref{f36}).

\section{Gaussian and quasi-Gaussian path integrals over velocities}

In the previous sections we demonstrated that all the propagators both in
nonrelativistic and relativistic quantum mechanics can be only presented
by means of bosonic path integrals over velocities of space-time coordinates.
All these integrals have the following structure
\begin{equation}\label{g1}
\int {\cal D}v \; \delta^{n}\left(\int v d\tau \right)F\left[v\right] ,
\end{equation}
\noindent where
\begin{eqnarray}\label{g2}
&&{\cal D}v=Dv\left[ \int Dv\;\;\delta ^n\left( \int v d\tau
 \right) \exp \left \{ i\int d\tau \left(-\frac{v^2}2\right)\right \}
\right] ^{-1}\;, \\
&&v=(v^{\mu}), \; v^{2}=\eta_{\mu\nu}v^{\mu}v^{\nu}, \; \mu,\nu=\overline
{0,n-1}, \nonumber
\end{eqnarray}
\noindent with some functional $F\left[v\right]$. In relativistic case
$n=4$, and  $\eta_{\mu\nu}$ is Minkowski tensor; in nonrelativistic case $n=3$,
and $\eta_{\mu\nu}$ is reducing to $-\delta_{ij}$.

Ways of doing path integrals of general form are unknown at present time,
only Gaussian path integrals, treated in certain sense, can be taken directly.
That is also valid with regards to the integrals in question (\ref{g1}).
However, if we restrict ourselves with a limited class
of functionals $F[v]$, which are called quasi-Gaussian
and are defined below, then one can formulate some
universal rules of their calculation and handling them. Similar idea has been
realized in the field theory \cite{b17,b18}. The restriction with
quasi-Gaussian functionals corresponds, in fact, to a perturbation theory,
with Gaussian path integral as a zero order approximation.

Introduce  Gaussian functional $F_{G}[v,I]$,
\begin{equation}\label{g3}
F_{G}[v,I]=\exp  \left\{-\frac{i}{2}\int d\tau d\tau^{\prime}
v^{\mu}(\tau)L_{\mu \nu}(g,\tau, \tau^{\prime})v^{\nu}(\tau^{\prime})
-i\int d\tau I_{\mu}(\tau)v^{\mu}(\tau) \right\}\;,
\end{equation}
\noindent where $v^{\mu}(\tau)$ are the velocities and $I_{\mu}(\tau)$ are
corresponding sources to them. A functional $F_{q G}[v,I]$ we
call  quasi-Gaussian if
\begin{equation}\label{g4}
F_{q G}[v,I]=F[v]F_{G}[v,I]\;,
\end{equation}

\noindent where $F[v]$ is a functional, which can be expanded in the functional
series of $v$,
\begin{equation}\label{g5}
F[v]=\sum_{n=0}\int d\tau_{1} \ldots d\tau_{n}
F_{\mu_{1} \ldots \mu_{n}}\left(\tau_{1} \ldots \tau_{n}\right)
v^{\mu_{1}}(\tau_{1})\ldots v^{\mu_{n}}(\tau_{n})\;.
\end{equation}

\noindent In (\ref{g3}) the matrix $L_{\mu \nu}(g,\tau, \tau^{\prime})$
supposes to have the following form
\begin{equation}\label{g6}
L_{\mu \nu}(q,\tau, \tau^{\prime})=\eta_{\mu \nu}\delta (\tau-\tau^{\prime})
+gM_{\mu \nu}(\tau, \tau^{\prime})\;.
\end{equation}

Define the path integral over velocities $v$ of the Gaussian functional
as
\begin{eqnarray}\label{g7}
& &\int {\cal D}v \;\delta^{n}\left(\int v d\tau \right) F_{G}[v,I] \\
& &=\left[ \frac{{\rm Det}\;L(g)\;\det\;l(g)}{{\rm Det}\;L(0)\;\det\;
l(0)}\right]^{-1/2}
\exp \left\{\frac{i}{2}\int d\tau d\tau^{\prime}
I(\tau)K(\tau, \tau^{\prime})I(\tau^{\prime}) \right \}\;,
\nonumber
\end{eqnarray}

\noindent where
\begin{eqnarray}\label{g8}
& &K(\tau, \tau^{\prime})=L^{-1}(g,\tau, \tau^{\prime})-Q^{T}(\tau)
l^{-1}(g)Q(\tau^{\prime}) \; , \nonumber \\
& &l(g)=\int d\tau d\tau^{\prime}L^{-1}(g,\tau, \tau^{\prime}),\;
\;Q(\tau)=\int d\tau^{\prime} L^{-1}(g,\tau^{\prime}, \tau) \; .
\end{eqnarray}

\noindent The formula (\ref{g7}) can be considered as infinitedimensional
generalization of the straightforward calculations result in the frame
of the discretization procedure, connected with the original definition of
path integrals for propagators discussed in the previous sections. In course
of doing of finitedimensional integrals it is implied a supplementary
definition of arisen improper Gaussian integrals by means of the analytical
continuation in the matrix elements of the nonsingular matrix $L$.

To avoid problems with calculations of determinants of matrices with
continuous indices we can use the formula

\begin{equation}\label{g9}
\frac{{\rm Det}\;L(g)}{{\rm Det}\;L(0)}=\exp \left\{
\int_{0}^{g} dg^{\prime}{\rm Tr}\;L^{-1}(g^{\prime})M \right\}\;,
\end{equation}

\noindent which may be derived similar to one (\ref{f25}). Taking into
account that $\det\; l(0)=-1$, we can rewrite the
path integral of the Gaussian functional in the following form
\begin{eqnarray}\label{g10}
& &\int {\cal D}v \;\delta^{n}\left(\int v d\tau \right) F_{G}[v,I] \\
& &=\left[ -\det\;l(g)\right]^{-1/2}
\exp \left \{ \frac{i}{2}\int d\tau d\tau^{\prime}I(\tau)K(\tau, \tau^{\prime})
I(\tau^{\prime})-\frac{1}{2}\int_{0}^{g} dg^{\prime}{\rm Tr}\;
L^{-1}(g^{\prime})M \right\}\;.\nonumber
\end{eqnarray}

The path integral of the quasi-Gaussian functional we define through one of
the Gaussian functional
\begin{eqnarray}\label{g11}
& &\int {\cal D}v \;\delta^{n}\left(\int v d\tau \right) F_{q G}[v,I]
=F\left(i\frac{\delta}{\delta I}\right)\int {\cal D}v \;\delta^{n}
\left(\int v d\tau \right) F_{G}[v,I]
\nonumber \\
& &=\left[- \det \;l(g)\right]^{-1/2}F \left(i\frac{\delta}{\delta I}\right)
\exp \left \{ \frac{i}{2} \int d\tau d\tau^{\prime}
I(\tau)K(\tau, \tau^{\prime})I(\tau^{\prime})-\frac{1}{2}\int_{0}^{g}
dg^{\prime}{\rm Tr}\;L^{-1}(g^{\prime})M \right\}\;.
\end{eqnarray}

One can formulate rules of handling  integrals from quasi-Gaussian functionals,
using the formula (\ref{g11}). For example, such integrals are
invariant under shifts of integration variables,
\begin{equation}\label{g12}
\int {\cal D}v \;\delta^{n}\left(\int (v+u) d\tau \right) F_{q G}[v+u,I]=
\int {\cal D}v\; \delta^{n}\left(\int v d\tau \right) F_{q G}[v,I]\; .
\end{equation}

\noindent The validity of this assertion for the Gaussian integral can be
verified by a direct calculation. Then the general formula (\ref{g12}) follows
from the (\ref{g11}). Using the property (\ref{g12}), one
can derive an useful generalization of the formula (\ref{g11}),
\begin{eqnarray}\label{g13}
& &\int {\cal D}v \;\delta^{n}\left(\int vd\tau -a\right) F_{qG}[v,I]
\nonumber\\
& &=\left[ -\det\;l(g)\right]^{-1/2}F \left(i\frac{\delta}{\delta I}\right)
\exp \left \{ \frac{i}{2}\int d\tau d\tau^{\prime}
I(\tau)K(\tau, \tau^{\prime})I(\tau^{\prime})\right.
\nonumber \\
& &\left.-\frac{i}{2}al^{-1}(g)a  -ial^{-1}(g)\int Q(\tau)I(\tau)d\tau
-\frac{1}{2}\int_{0}^{g} dg^{\prime}{\rm Tr}\;
L^{-1}(g^{\prime})M  \right \}\;,
\end{eqnarray}

\noindent where $a$ ia a constant vector. The integral of the total
functional derivative over $v^{\mu}(\tau)$ is equal to zero,
\begin{equation}\label{g14}
\int {\cal D}v \;\frac{\delta}{\delta v^{\mu}(\tau)}
\;\delta^{n}\left(\int v d\tau \right) F_{q G}[v,I]=0\;.
\end{equation}
This property may be obtained as a consequence of the functional integral
invariance under the shift of variables, as well as by direct calculations of
integral (\ref{g14}). Using the latter, one
can derive formulas of integration by parts, which we do not present here. If
a quasi-Gaussian functional depends on a parameter $\alpha\;,$ then
the derivative with respect to this parameter is commutative with the integral
sign,
\begin{equation}\label{g15}
\frac{\partial}{\partial \alpha}\int {\cal D}v\; \delta^{n}
\left(\int v d\tau \right) F_{q G}[v,I,\alpha]
=\int {\cal D}v \;\delta^{n}\left(\int v d\tau \right) \frac{\partial}{
\partial \alpha}F_{q G}[v,I,\alpha] \;.
\end{equation}

\noindent Finally, the formula for the change of the variables holds:
\begin{equation}\label{g16}
\int {\cal D}v \;\delta^{n}\left(\int v d\tau \right) F_{q G}[v,I]
=\int {\cal D}v \;\delta^{n}\left(\int \phi d\tau \right)
F_{q G}[\phi,I] \;{\rm Det}\;\frac{\delta \phi_{\tau}(v)}
{\delta v(\tau^{\prime})}\; ,
\end{equation}

\noindent where $\phi_{\tau}(v)$ is a set of analytical functionals in $v\;,$
parameterized by $\tau\;.$
One can prove formulas (\ref{g15},\ref{g16}) in the same manner it was done
in \cite{b18} for the case of the field theory.

Thus, in  quantum mechanics, in the frame of perturbation theory,
one can define quasi-Gaussian path integrals over velocities and rules of
handling them. This definition is  close to one in field theory
\cite{b17,b18}, the analogy is stressed by the circumstance that, as in the
field theory, the  integrals over velocities do not contain explicitly any
boundary condition for trajectories of the integration. After the rules of
integration are formulated, one can forget about the origin of the integrals
over velocities and fulfil integrations, using the rules only. In the next
Section we demonstrate this technique on some examples.

\section{Example}

Here we are going to  calculate the propagator of a scalar particle
in an external electromagnetic field, using representation (\ref{f8})
and rules of integrations, presented in the previous sections.
We consider a combination of a constant
homogeneous field and a plane wave field. The potentials for this
field may be taken in the form
\begin{equation}
\label{h1}
A_\mu \left( x\right) =-\frac 12F_{\mu \nu }x^\nu +f_\mu \left(
nx\right)\;,
\end{equation}

\noindent where $F_{\mu \nu}$ is the field strength tensor of the constant
homogeneous field with nonzero invariants
\begin{eqnarray*}
{\cal F}&=&\frac14 F_{\mu \nu }F^{\mu \nu }\neq 0\;,\;
\;{\cal G}=-\frac14 F^{\ast}_{\mu \nu }F^{\mu \nu }\neq 0\;,
\end{eqnarray*}

\noindent $(F^{\ast}_{\mu \nu }=\frac{1}{2}\epsilon_{\mu \nu \alpha \beta}
F^{\alpha \beta},\;\epsilon_{\mu \nu \alpha \beta}$
is totally antisymmetric
tensor), in terms of which its eigenvalues ${\cal E}$ and ${\cal H}$
are expressed
\begin{eqnarray}
\label{h2}
&&F_{\mu \nu }n^\nu =-{\cal E}n_\mu\;,\;\; F_{\mu \nu }\bar n^\nu ={\cal E}
\bar n_\mu \;, \;\;
F_{\mu \nu }\ell^\nu=i{\cal H}\ell_{\mu}\;,\;\;
F_{\mu \nu }\bar {\ell}^\nu =-i{\cal H}
\bar{\ell}_\mu \;, \\
&&{\cal E}=\left[ ( {\cal F}^2+{\cal G}^2) ^{\frac 12}-
{\cal F}\right] ^{\frac 12}\;,\;{\cal H}=\left[ ( {\cal F}^2+
{\cal G}^2) ^{\frac 12}+{\cal F}\right)]^{\frac 12}\;.
\nonumber
\end{eqnarray}
\noindent The eigenvectors $n,\;\bar{n},\;
\ell,\;\bar{\ell}$ are isotropic and obey the conditions
\begin{equation}
\label{h3}
n^2 =\bar{n}^2 =\ell ^2 =\bar{\ell}^2 =0\;,
\;\; n \bar{n}=2\;,\;\ell \bar{\ell}=-2\;,\;
n\ell=\bar{n} \ell=n \bar{\ell}=\bar{n} \bar{\ell}=0\;.
\end{equation}

\noindent The functions  $f_\mu \left( nx\right) $
are arbitrary, except for the fact that they are subject to the conditions
\begin{equation}
\label{h4}
f_\mu \left( nx\right) n^\mu =f_\mu \left( nx\right)
\bar n^\mu =0 \;.
\end{equation}

\noindent The total field strength tensor for the potential (\ref{h1})
is
\begin{equation}
\label{h5}
F_{\mu \nu }( x) =F_{\mu \nu }+\Psi _{\mu \nu }(
nx),\; \Psi _{\mu \nu }( nx) =n_\mu f_\nu
^{\prime }( n x) -n_\nu f_\mu ^{\prime }
( nx) \;.
\end{equation}

\noindent Since the invariants ${\cal F}\;,\;{\cal G}$ of the tensor
$F_{\mu \nu}$ are nonzero, there exists a special reference frame, where
the electric and magnetic fields, corresponding to this tensor, are collinear
with respect to one another and to the spatial part ${\bf n}$ of the
four-vector $n$.
In this reference frame, the total  field $F_{\mu \nu}(x)$ corresponds
to a constant homogeneous and collinear electric and magnetic
fields together with a plane wave, propagating along them;
${\cal E}\;,{\cal H}$, being equal to the strengths of a constant homogeneous
electric and
magnetic fields, respectively. In terms of the defined eigenvectors
the tensor $F_{\mu \nu}$ can be written as
\begin{equation}
F_{\mu \nu}=\frac{{\cal E}}{2}\left(\bar{n}_{\mu}n_{\nu}-
n_{\mu}\bar{n}_{\nu}\right)+\frac{i{\cal H}}{2}\left(\bar{\ell}_{\mu}
\ell_{\nu}-\ell_{\mu}\bar{\ell}_{\nu}\right)\;,
\label{h6}
\end{equation}

\noindent and the completeness relation holds
\begin{equation}
\eta_{\mu \nu}=\frac{1}{2}\left( \bar{n}_{\mu}n_{\nu}+
n_{\mu}\bar{n}_{\nu}-\bar{\ell}_{\mu}
\ell_{\nu}-\ell_{\mu}\bar{\ell}_{\nu}\right)\;.
\label{h7}
\end{equation}

\noindent The latter allows one to express any
four-vector $u$ in terms of the eigenvectors (\ref{h2}),
\begin{eqnarray}
\label{h8}
& &u^{\mu}=n^{\mu}u^{(1)} +\bar{n}^{\mu}u^{(2)} +\ell^{\mu}u^{(3)}
 +\bar{\ell}^{\mu}u^{(4)}\;,
\nonumber \\
& &u^{(1)}=\frac{1}{2}\bar{n}u\;,\;u^{(2)}=\frac{1}{2}nu\;,\;
u^{(3)}=-\frac{1}{2}\bar{\ell}u\;,\;u^{(4)}=-\frac{1}{2}\ell u\;.
\end{eqnarray}

In these concrete calculations it is convenient for us
to make a shift of variables in the formula
(\ref{f9}), to
rewrite it in the following form
\begin{eqnarray}
& &\Delta(e_0)=\exp \left ( i\frac{ \Delta x^2 }{2e_0}\right )
\int {\cal D}v\;\delta ^4\left( \int v d\tau  -
\frac{\Delta x}{\sqrt{e_0}}\right ) \nonumber \\
& &\times \exp \left\{ i\int d\tau \left[ -\frac{v^2}2-g\sqrt{e_0}v
A\left( \sqrt{e_0}\int_{0}^{\tau}
v( \tau ^{\prime })d\tau ^{\prime } +x_{in} \right) \right] \right \} \; .
\label{h9}
\end{eqnarray}

The calculations will be made in two steps: first in a constant
homogeneous field only,
and then in the total combination (\ref{h1}), using some results of the first
problem.
Thus, on the first step the potentials of the electromagnetic field
are
\begin{equation}
\label{h10}
A_\mu \left( x\right) =-\frac 12F_{\mu \nu }x^\nu \;.
\end{equation}

\noindent Substituting the external field (\ref{h10}) into (\ref{h9}), one
can find
\begin{eqnarray}\label{h11}
& &\Delta(e_0)=\exp \left(i\frac{\Delta x^2}{2e_0}\right)\int {\cal D}v\;
\delta^4\left(\int vd\tau -\frac{\Delta x}{\sqrt{e_0}}\right)\\
& &\times \exp \left\{-\frac{i}{2}\int d\tau d \tau^{\prime} v(\tau)
L\left(g,\tau,\tau^{\prime}\right)v(\tau^{\prime})
-i\int \frac{g\sqrt{e_{0}}}{2}\;x_{in}F\;vd\tau \right\}\;,
\nonumber
\end{eqnarray}

\noindent where
\begin{equation}
L_{\mu \nu }(g, \tau ,\tau ^{\prime })=\eta _{\mu \nu }\delta
( \tau -\tau ^{\prime }) -\frac{ge_{0}}{2}F_{\mu \nu }\epsilon
(\tau -\tau ^{\prime })\; .
\label{h12}
\end{equation}

\noindent The path integral
(\ref{h11}) is the Gaussian one (see (\ref{g13})).
To get an answer, one needs to find the inverse matrix
$L^{-1}(g, \tau ,\tau ^{\prime })$, which satisfies
the equation

\[
\int L(g, \tau ,\tau ^{\prime \prime})
L^{-1}(g, \tau^{\prime \prime} ,\tau ^{\prime })d\tau^{\prime \prime}=
\delta(\tau-\tau^{\prime})\;.
\]

\noindent One can demonstrate, that this equation is equivalent to a
differential one,
\begin{equation}
\label{h13}
\frac{\partial}{\partial \tau}L^{-1}(g, \tau ,\tau ^{\prime})
-ge_{0}F\;L^{-1}(g, \tau ,\tau ^{\prime})=\delta^{\prime}
(\tau-\tau^{\prime})\;,
\end{equation}

\noindent with initial condition

\[
L^{-1}(g, 0,\tau ^{\prime})+\frac{ge_{0}F}{2}
\int L^{-1}(g, \tau^{\prime \prime} ,\tau ^{\prime})d\tau^{\prime \prime}
=\delta(\tau^{\prime})\;.
\]
\noindent Its solution has the form
\begin{equation}\label{h14}
L^{-1}(g, \tau ,\tau ^{\prime })
=\delta ( \tau -\tau^{\prime })
+\frac{ge_{0}F}{2}\exp \left\{ ge_{0}(\tau -\tau ^{\prime})F
\right\}
\left[ \epsilon(\tau -\tau ^{\prime }) -
\tanh \left(\frac{ge_{0}F}{2}\right) \right]\;.
\end{equation}

\noindent Using (\ref{h14}), one can find all ingredients of the general
formula (\ref{g13}), taking into account that
$a=-\Delta x/\sqrt e_{0}\;,\;\; I(\tau)=g\sqrt{e_{0}}
x_{in}F/2\;.$

\noindent Thus,
\begin{eqnarray}\label{h15}
K( \tau ,\tau ^{\prime })
& &=\delta ( \tau -\tau^{\prime })
+\frac{ge_{0}F}{2}\exp \left\{ ge_{0}(\tau -\tau ^{\prime})F
\right\}
\left[ \epsilon(\tau -\tau ^{\prime }) -
\coth \left(\frac{ge_{0}F}{2}\right) \right]\;,\\
& &\int d\tau d\tau^{\prime}K(\tau ,\tau^{\prime})=0 \;,
\; \int d\tau Q(\tau)=l(g) \;,\;
l(g)=\frac{\tanh ge_{0}F/2}{ge_{0}F/2}\;,
\nonumber \\
& &M(\tau,\tau^{\prime})=-\frac{e_0}{2}F \epsilon(\tau -\tau ^{\prime })\;,
\;
\int_{0}^{g} dg^{\prime} {\rm Tr}L^{-1}(g^{\prime})M=
{\rm tr}\ln (\cosh ge_{0}F/2) \;,
\nonumber
\end{eqnarray}

\noindent where the symbol ``${\rm tr}$'' is being taken over four dimensional
indices only. Then
\begin{eqnarray}
\label{h16}
& &\Delta \left( e_0\right)=
\left[- \det \left(\frac{\sinh ge_0F/2}{gF/2}\right)\right]^{-1/2}\\
& &\times \exp \left\{ \frac{i}{2}\left[\frac{\Delta x^2}{e_0}
+ g\;x_{out}Fx_{in}-\frac{1}{2}\Delta x \;gF\coth
\left(\frac{ge_0F}2\right)\Delta x\right]\right\}\;.
\nonumber
\end{eqnarray}

\noindent Substituting (\ref{h16}) into (\ref{f8}), we get a final
expression for the causal propagator of a scalar particle
in a constant homogeneous  electromagnetic field
\begin{eqnarray}
\label{h17}
& &D^c\left( x_{out},x_{in}\right)=\frac 1 {2\left( 2\pi \right)
^2}\int_0^\infty de_o\left[- \det \left(\frac{\sinh ge_0F/2}{gF/2}
\right)\right]^{-1/2}\\
& &\times \exp \left\{ \frac {i}{2}\left[gx_{out}Fx_{in}-e_{0}m^2-\frac{1}{2}
\Delta x\;gF\coth (\frac{ge_0F}2)\Delta x\right]\right\}\;.
\nonumber
\end{eqnarray}

\noindent This result was first derived by Schwinger, using his proper time
method \cite{b27}.

Now we return to the total electromagnetic field (\ref{h1}).
Let us substitute the potential (\ref{h1}) into (\ref{h9}),
\begin{eqnarray}
\label{h18}
& &\Delta(e_0)=\exp \left(i\frac{\Delta x^2}{2e_0}\right)\int {\cal D}v\;
\delta^4\left(\int vd\tau -\frac{\Delta x}{\sqrt{e_0}}\right)\\
& &\times \exp \left\{-\frac{i}{2}\int d\tau d \tau^{\prime} v(\tau)
L\left(g,\tau,\tau^{\prime}\right)v(\tau^{\prime})-
i\int \frac{g\sqrt e_{0}}{2}\;x_{in}F\;vd\tau \right.
\nonumber \\
& &\left.-ig\sqrt e_0 \int d\tau v(\tau) f\left(n x_{in}+\sqrt e_0
\int_0^\tau n v(\tau^{\prime})d\tau^{\prime}\right)\right\}\;,
\nonumber
\end{eqnarray}

\noindent with $L(g,\tau,\tau^{\prime})$  defined in (\ref{h12}).
One can take the integral (\ref{h18}) as quasi-Gaussian, in accordance
with the formula (\ref{g13}). So, one can write
\begin{eqnarray}
\label{h19}
& &\Delta(e_0)\\
&=&\exp \left\{g\sqrt e_0 \int d\tau f\left(nx_{in}+i\sqrt e_0
\int_0^\tau n\frac{\delta}{\delta I(\tau^{\prime})}
d\tau^{\prime}\right)\frac{\delta}{\delta I(\tau)}\right\}B\left(I\right)|_
{I=0}\;,\nonumber
\end{eqnarray}

\noindent where
\begin{eqnarray}
\label{h20}
& &B\left(I\right)=\exp \left(i\frac{\Delta x^2}{2e_0}\right)\int {\cal D}v\;
\delta^4\left(\int vd\tau -\frac{\Delta x}{\sqrt{e_0}}\right)\\
& &\times \exp \left\{-\frac{i}{2}\int d\tau d \tau^{\prime} v(\tau)
L\left(g,\tau,\tau^{\prime}\right)v(\tau^{\prime})
-i\int \left(\frac{g\sqrt{e_0}}{2}x_{in}F+I(\tau)\right)v(\tau)
d\tau\right \} \;.
\nonumber
\end{eqnarray}

\noindent The integral can be found similar to
(\ref{h11}). As a result we get
\begin{equation}\label{h21}
B\left(I\right)=\exp \left\{\frac{i}{2}
\int d\tau d \tau^{\prime} I(\tau)
K\left(\tau,\tau^{\prime}\right)I(\tau^{\prime})-
i\int I(\tau)a(\tau) d\tau \right \}
\Delta(e_{0})|_{\Psi=0} \;,
\end{equation}

\noindent where $\Delta(e_{0})|_{\Psi=0}$ is the expression given
by (\ref{h16}),
$\;K(\tau,\tau^{\prime})$ is defined in (\ref{h15}), and

\[
a(\tau)=\frac{\Delta x}{2\sqrt e_0}\left(1+\coth\left(ge_{0}F/2\right)
\right)ge_{0}F\exp\left(-ge_{0}F\tau \right)\;.
\]

To obtain the action of the operator, involved in (\ref{h19}),
on the functional $B\left(I\right)$, we decompose the sources $I^{\mu}(\tau)$
in the eigenvectors (\ref{h2}), using (\ref{h8})

\[
I^{\mu}(\tau)=\frac{1}{2}\left(n^{\mu}\;\bar{n}I(\tau) +\bar{n}^{\mu}\;nI(\tau)
-\ell^{\mu}\;\bar{\ell}I(\tau) -\bar{\ell}^{\mu}\;\ell I(\tau) \right)\;.
\]

\noindent Then, it is possible to write
\[
n\frac{\delta}{\delta I(\tau)}=2\frac{\delta}{\delta \bar{n}I(\tau)}\;,\;\;
f\frac{\delta}{\delta I(\tau)} = \bar{\ell} f
\frac{\delta}{\delta \bar{\ell} I(\tau)}
+\ell f\frac{\delta}{\delta \ell I(\tau)}\;.
\]

\noindent Using this, we get
\begin{eqnarray*}
& &f\left(nx_{in}+i\sqrt e_0\int_0^\tau n\frac{\delta}
{\delta I(\tau^{\prime})}d\tau^{\prime}\right)\frac{\delta}{\delta I(\tau)} \\
& &= \bar{\ell} f\left(nx_{in}+i\sqrt e_0\int_0^\tau \frac{\delta}
{\delta \bar{n}I(\tau^{\prime})}d\tau^{\prime}\right)
\frac{\delta}{\delta \bar{\ell} I(\tau)}+\ell f\left(nx_{in}+i\sqrt e_0
\int_0^\tau \frac{\delta}{\delta \bar{n}I(\tau^{\prime})}
d\tau^{\prime}\right)\frac{\delta}{\delta \ell I(\tau)}\;,\\
& &\int d\tau d\tau^{\prime}I(\tau)K\left(\tau,\tau^{\prime}\right)
I(\tau^{\prime})\\
& &=\int d\tau d\tau^{\prime}\left[
\bar{n}I(\tau)\;nI(\tau^{\prime})K\left(\tau,\tau^{\prime},{\cal E}\right)
-\bar{\ell}I(\tau)\;\ell I(\tau^{\prime})
K\left(\tau,\tau^{\prime},i{\cal H}\right)\right]\;,\\
& &\int I(\tau)a(\tau)d\tau=\frac{1}{2}\int \left[\bar{n}I(\tau)\;na(\tau)
+nI(\tau)\;\bar{n}a(\tau)
- \bar{\ell}I(\tau)\;\ell a(\tau)-\ell I(\tau)\;\bar{\ell}a(\tau)
\right]d\tau\; ,
\end{eqnarray*}

\noindent where
\begin{eqnarray*}
& &K\left(\tau,\tau^{\prime},{\cal E}\right)=\delta ( \tau -\tau^{\prime })
+\frac{ge_{0}{\cal E}}{2}\exp \left\{ ge_{0}(\tau -\tau ^{\prime}){\cal E}
\right\}
\left[ \epsilon(\tau -\tau ^{\prime }) -
\coth \left(\frac{ge_{0}{\cal E}}{2}\right) \right]\;,\\
& &K\left(\tau,\tau^{\prime},i{\cal H}\right)=\delta ( \tau -\tau^{\prime })
+\frac{ige_{0}{\cal H}}{2}\exp \left\{ ige_{0}(\tau -\tau ^{\prime}){\cal H}
\right\}
\left[ \epsilon(\tau -\tau ^{\prime }) +i
\cot \left(\frac{ge_{0}{\cal H}}{2}\right) \right] \;.
\end{eqnarray*}

\noindent Now the exponent of the functional $B[I]$ is linear in
$nI(\tau)\;,\;\bar{n}I(\tau)\;,\;\ell I(\tau)\;,\\
\bar{\ell}I(\tau).\;$ Thus, one can easy to get a result
\begin{eqnarray}
& &\Delta(e_{0})=\exp \left\{\frac{i}{2}
g^{2}e_{0}
\int d\tau d\tau^{\prime} f\left(nx_{cl}(\tau)\right)
K(\tau, \tau^{\prime})f\left(nx_{cl}(\tau^{\prime})\right)\right.
\nonumber \\
& &\left.+ig\sqrt e_{0} \int d\tau a(\tau)f\left(nx_{cl}(\tau)\right)\right\}
\Delta(e_{0})|_{\Psi=0}\;,
\label{h22}\\
& &nx_{cl}(\tau)=nx_{in}+\frac{1-\exp\left(ge_{0}{\cal E}\tau\right)}
{1-\exp\left(ge_{0}{\cal E}\right)}n\Delta x \;,
\;\; nx_{cl}(0)=nx_{in}\;,\;\;nx_{cl}(1)=nx_{out}\;,
\nonumber
\end{eqnarray}

\noindent where $x_{cl}(\tau)$ is the solution \cite{b12} of the Lorentz
equation in the external electromagnetic field (\ref{h1}).
Gathering ({\ref{h22}) and (\ref{h16}), we get
\begin{eqnarray}
& &\Delta\left(e_0\right)=\left[-\det \frac{\sinh \left(ge_{0}F/2\right)}
{ge_{0}F/2}\right]^{-1/2}\exp\left\{\frac{i}{2}\left[gx_{out}Fx_{in}
\right.\right.\nonumber \\
& &\left.\left.-\frac{1}{2}\left(\Delta x+l(e_{0},1)\right)\;gF\coth \left(
ge_{0}F/2\right)\;\left(\Delta x+l(e_{0},1)\right)+2\Phi(e_{0})\right.\right.
\nonumber \\
& &\left.\left.+ \Delta x\;gFl(e_{0},1)+
\frac{\Delta x^2}{e_{0}}\right] \right\}\;,
\label{h23}
\end{eqnarray}

\noindent where
\begin{eqnarray}
\label{h24}
\Phi(e_{0})&=&e_{0}\int g f\left(nx_{cl}(\tau)\right)\left[
gf\left(nx_{cl}(\tau)\right) +gFl(e_{0},\tau)\right]d\tau\;,\\
l(e_{0},\tau)&=&e_{0}\int_{0}^{\tau}\exp\left\{ge_{0}(\tau-\tau^{\prime})F
\right)gf\left(nx_{cl}(\tau^{\prime})\right\} d\tau^{\prime}\;.
\nonumber
\end{eqnarray}

\noindent Substituting (\ref{h23}) into (\ref{f8}), we arrive to the final
expression
for the causal propagator of a scalar particle in the external
electromagnetic field (\ref{h1}):
\begin{eqnarray}
\label{h25}
& &D^c\left( x_{out},x_{in}\right)=\frac 1 {2\left( 2\pi \right)
^2}\int_0^\infty de_{0}\left[- \det \left(\frac{\sinh ge_0F/2}{gF/2}
\right)\right]^{-1/2}\\
& &\times \exp \left\{ \frac {i}{2}\left[gx_{out}Fx_{in}-e_{0}m^2
+\Delta x\;gFl(e_{0},1) \right.\right.
\nonumber \\
& &\left.\left.-\frac{1}{2}\left(\Delta x+l(e_{0},1)\right)\;gF\coth \left(
ge_{0}F/2\right)\;\left(\Delta x+l(e_{0},1)\right)+2\Phi(e_{0})
\right]\right\}\;.
\nonumber
\end{eqnarray}

\noindent This expression coincides with the one,
obtained  in
\cite{b28}, by means of the method of summation over
exact solutions of Klein-Gordon equation in the
external field (\ref{h1}). A detailed description of quantum electrodynamical
processes in such a field one can find in \cite{b12,b29}.

\end{document}